\documentclass[%
 aip,
 amsmath,amssymb,
 reprint,%
]{revtex4-1}

\usepackage{graphicx}% Include figure files
\usepackage{dcolumn}% Align table columns on decimal point
\usepackage{bm}% bold math
\usepackage[utf8]{inputenc}
\usepackage[T1]{fontenc}
\usepackage{mathptmx}
\usepackage{etoolbox}

\makeatletter
\def\@email#1#2{%
 \endgroup
 \patchcmd{\titleblock@produce}
  {\frontmatter@RRAPformat}
  {\frontmatter@RRAPformat{\produce@RRAP{*#1\href{mailto:#2}{#2}}}\frontmatter@RRAPformat}
  {}{}
}%

\makeatother
\begin{document}

\preprint{AIP/123-QED}

\title[ERICA Methods]{Controlled Partial Gravity Platform for Milligravity in Drop Tower Experiments}

%\author{Kolja Joeris, Matthias Keulen, Jonathan E. Kollmer }
\author{Kolja Joeris}
\altaffiliation{University of Duisburg-Essen, Faculty of Physics, Lotharstr. 1, 47057 Duisburg, Germany}
\author{Matthias Keulen}
\affiliation{University of Duisburg-Essen, Faculty of Physics, Lotharstr. 1, 47057 Duisburg, Germany}
\author{Jonathan E. Kollmer}
\affiliation{University of Duisburg-Essen, Faculty of Physics, Lotharstr. 1, 47057 Duisburg, Germany}

\date{\today}

\begin{abstract}
We detail a platform for partial g environment and an experiment for simulated impacts on asteroid surfaces based on it. The partial g environment is created by a two stage approach: First, create microgravity using the ZARM drop tower. Second, convert microgravity to partial gravity by steady acceleration of experiment volume on linear drive inside microgravity environment. The experiment we conducted on this platform simulates low-velocity impacts into a simulated asteroid surface. To recreate the asteroid environment, in addition to the partial gravity, a vacuum chamber is used. We explain requirements, setup and operation of partial gravity platform and experiment and discuss its performance. Finally, we are open for requests for external experiments which might benefit from our platform with $9.3\,$s of controlled partial gravity down to the mm/s$^2$ range with low g-jitter.
\end{abstract}

\maketitle

\section{Introduction}
Partial gravity experiments are becoming increasingly important for granular matter research \citep{sanchez2023effects}. In the light of asteroid missions, \citep{daly2023successful,Fujiwara, lauretta2017osiris}, questions arised that need a broader data base which needs to be established by less complex and less expensive means than by sending probes to asteroids. Sorting effects and overall structure, internal and superficial, of asteroids remain objects of active research \citep{Shinbrot, michel2020collisional,joeris2022influence, barnouin2019shape}.  Partial gravity not only is important for simulation of asteroid environments in the scope of basic research, but also from a technical perspective. Touchdown missions and sample return to and from loosely bound rubble pile asteroids remain a challenging task \citep{ballouz2021modified, walsh2022near,yano2006touchdown,lauretta2021osiris}. Predicting granular dynamics for the target environment is crucial for mission success. Furthermore reduced gravity levels are a fundamental prerequisite for experiments that aim to answer questions in the scope of planet formation \citep{Steinpilz20} and the early solar system, with open questions on how granular material is accreted into planetesimals and abraded on objects in the protoplanetary disc \citep{wurm2021understanding}. With the newly fueled race to the moon, the need for understanding granular dynamics under reduced gravity conditions increases \citep{creech2022artemis}. Similarly, missions to mars \citep{farley2020mars} can yield more questions which can be investigated on earth, or benefit from ground based preliminary investigation.
\textbf{Platforms Overview:} There are currently a number of platforms for reduced gravity research. Each coming with its specific challenges and shortcomings. Parabolic flights can produce partial gravity, and are typically available for lunar and martian g-levels. In case of the European Zero-G Aircraft the total experiment time per flight day consists of  $31$ parabolas with approximately $20\,$s of reduced gravity per parabola. The acceleration profile, as seen in \citep{carr2018acceleration}, however are not suitable for highly sensitive experiments which are susceptible for perturbations in the mm/s$^2$ regime. Sounding rockets and the international space station as a platform provide exceptionally high microgravity times. Residual accelerations on the ISS and sounding rockets can be in the range of $10^{-4}\,$g, limiting its use for experiments requiring precise milligravity environments \citep{barmatz2007critical,steinpilz2019arise,thomas2000microgravity, Bila_2024}.  Compared to parabolic flights, ISS and sounding rockets offer limited space, massively increased cost and serious design restrictions with regard to experiment dimensions, power, interactivity and possible disturbances on other experiments, e.g. vibrations. In this set of platforms, the parabolic flights are the only ones providing partial g environments without implementation of further measures, which will place further constraints on the quality of the g levels provided. For rockets, space stations and satellites, typicall systems like centrifuges are used to convert mircrogravity to the targeted gravity levels  \citep{ozaki2023granular, musiolik2018saltation}. Centrifuges come with some inherent disadvantages again, with possible vibrations from the motor and bearings, and, as a systematic problems,  a radial gravity gradient as well as Coriolis forces affecting the trajectories of particles moving trough the experiment volume. Another way of simulating partial gravity is density matching, by suspending the specimen in a fluid or dense gas. While this may be suitable for larger objects or fluids \citep{beysens1988phase}, the behaviour of granular systems depends strongly on the surrounding medium \citep{mitarai2006wet} and charge states on the surface of particles cannot be generalized from vacuum to liquid media. The last example for microgravity and partial gravity platforms are drop towers. The microgravity on those, e.g. the ZARM drop tower at Bremen, can be extremely clean with a jitter of less than $10^{-6}$ g and offering more than $9\,$s experiment time \citep{selig2010drop, von2006new}, repetition rate however is limited to two experiments per day due to the need to fully evacuate the drop tower to avoid air drag. For a higher repetition rate, a new class of actily driven "drop"-towers have recently become available, such as the Einstein Elevator and the ZARM GTB Pro. The latter can provide $\approx 2.4\,s$ of microgravity \citep{gierse2017fast} or lunar gravity with serveral tens of launches per day. The Einstein Elevator at HITec  \citep{lotz2017einstein} is planned to operate in a similar niche, offering different partial g levels.

As is, none of the systems described above are suitable to conduct experiments under asteroid g levels, i.e. $10^{-2} - 10^{-4} g$. The only platform currently available with a jitter significantly less than this desired g-level is the ZARM Bremen droptower. So, in order to access controlled millli-gravity conditions, we have devised a partial-g add-on for the ZARM Bremen droptower capsules, which we will detail in this manuscript. 

\textbf{Controlled milligravity platform :} Our system adds a controlled high precision linear stage to the ZARM droptower capsules. The function of the linear stage is to provide constant acceleration, while the drop tower capsule is in free fall.  To allow for asteroid level gravities the system needs to be able to provide accelerations down to the range of mm/s$^2$ whith a jitter level siginificantly lower. The gravity level should be adjustable and for experiment preparation time resolved acceleration profiles need to be available as well. The idea is that the experiment should be able to start at higher gravity levels and ease into the planned target g-value. The choice of a linear stage (in contrast to a centrifuge) is motivated by the aim to provide a clean milligravity environment without (noticable) vibrations, g-gradients and coriolis force. For the specific use case of experiments on the surfaces of rubble pile asteroids the experimet environment should also provide a vacuum. For easy experiment preparation the experiment chamber should be esaily accessible to be replaced or reset between experiment. To study impacts with velocities comparable to the escape velocities of smaller rubble pile asteroids, a controlled low velocity launcher that can inject material with velocities in the cm/s range, into the experimental chamber is necessary. 

%Partial gravity experimente zunehmend wichtig, beispeiele: Asteroiden, Planetentstehungen, neue Mondwettrennen, kommerzialisierung bla bla bla .. mit Verweis auf das NASA Whitepaper von Paul Sanchez.

%Bisherige Optionen für partial g und warum sie für manche experimente nicht geeignet sind (parabelflug -> zuviel jitter, zentrifuge hat corioliskraft (und/oder vibration, siehe japanerpaper), density matching is für staubdynamik blöd, .

%=> es gibt bedarf der bisher nicht gedeckt ist Jetzt kommen wir und öffnen eine neue nische mit dem hier vorgestellten system. 

%2 stufig, linear, am fallturm. bla blba....

%ziel g level von bis ... experimentzeit bis 10 sekunden ...blabla requirements ...

\section{Systems Overview}
\begin{figure*}
    \centering
    \includegraphics[width=\textwidth]{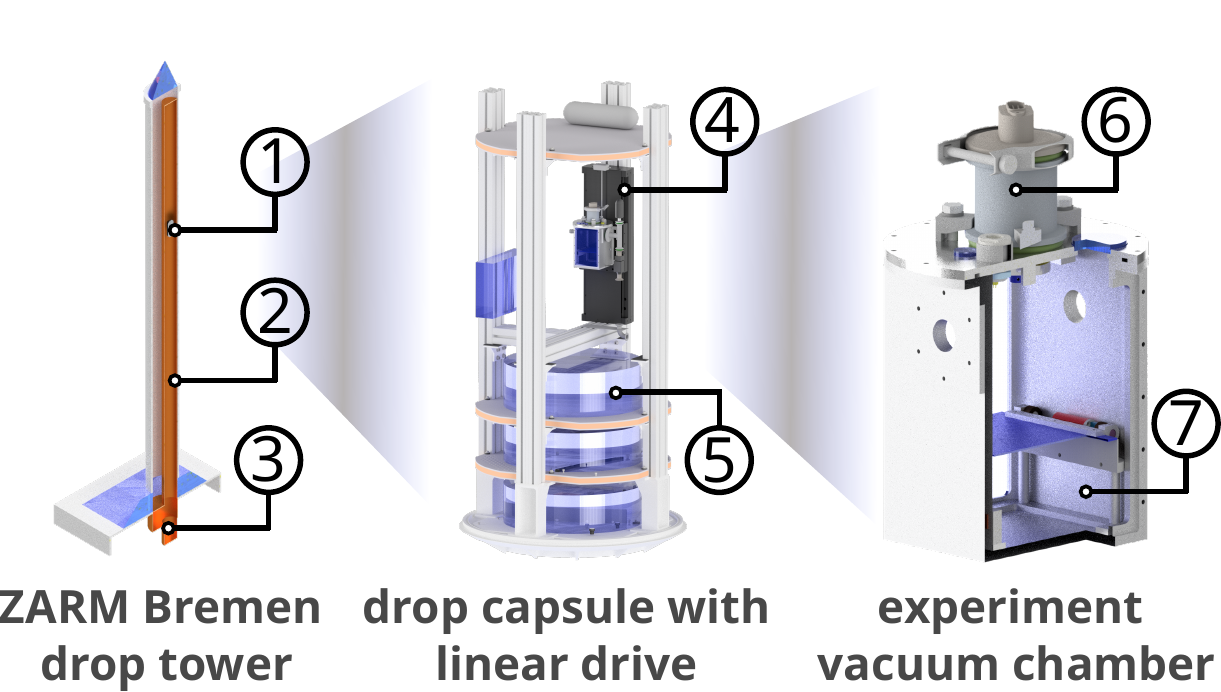}
    \caption{Two-stage process to create a precisely controlled partial gravity environment. Left panel: The ZARM Bremen drop tower provides a zero gravity environment (image credit: ZARM). It contains a vacuum tube (2), with a catapult (3) at the bottom. The capsule (1) falls freely inside of the vacuum tube. Center panel: The free falling drop-capsule contains a linear driving unit (1) which accelerated the experiment cell to create partial gravity. The drive as well as the experiment cell are controlled by a shock proof (for landing) computer (5). The drop capsule also contains vacuum- and battery systems (not shown). Right panel: The experiment cell is a vacuum chamber with granular bed compartment (7) and launcher module (6). }
    %The right panel shows a view of the high speed camera into the experiment cell during a low velocity (10 cm/s) impact experiment into regolith under asteroid gravity ($10^{-3}\,$g).
    \label{fig:vgg}
\end{figure*}

Our gravity control and asteroid environment simulation system is based on a two stage design: Microgravity is provided by the ZARM drop tower \citep{dropmanual}. The drop tower consists of a vacuum tube with a height of $120\,$m and a diameter of $3.5\,$m. It is surrounded by the supporting structure. The vacuum is needed to remove residual acceleration caused by air drag, when the drop tower capsule is in free fall inside of the tower. While the full height of the tower is only sufficient for $4.74\,$s of free fall, the time is almost doubled to $9.3\,$s by using a catapult at ground level to launch the capsule \citep{von2006new}. A deceleration container filled with polystyrene granules to a height of $8.2\,$m is positioned to catch the drop tower capsule upon landing. While the drop tower itself is evacuated, the interior of the drop capsule, the assembly that is being launched or dropped inside of the tower, remains pressurized to $\approx 1\,$bar. With a total mass of up to $500\,$kg it houses the experiments and, in this case, the second stage of our milligravity generation system: A linear stage, capable of accelerating a given object at a constant and precisely controllable rate, thus subjecting a sample container to a defined partial gravity level. Specifically, for the inital experiments referenced in this manuscript the linear stage carries a vacuum chamber with a set of cameras, lighting, specimen retainment system and impactor launcher. An overview of the complete two stage system and the experiment chamber can be seen in Fig. \ref{fig:vgg}.
\section{Individual Systems}
\subsection{Capsule}

\begin{figure}[ht]
    \centering
    \includegraphics[width=0.5\textwidth]{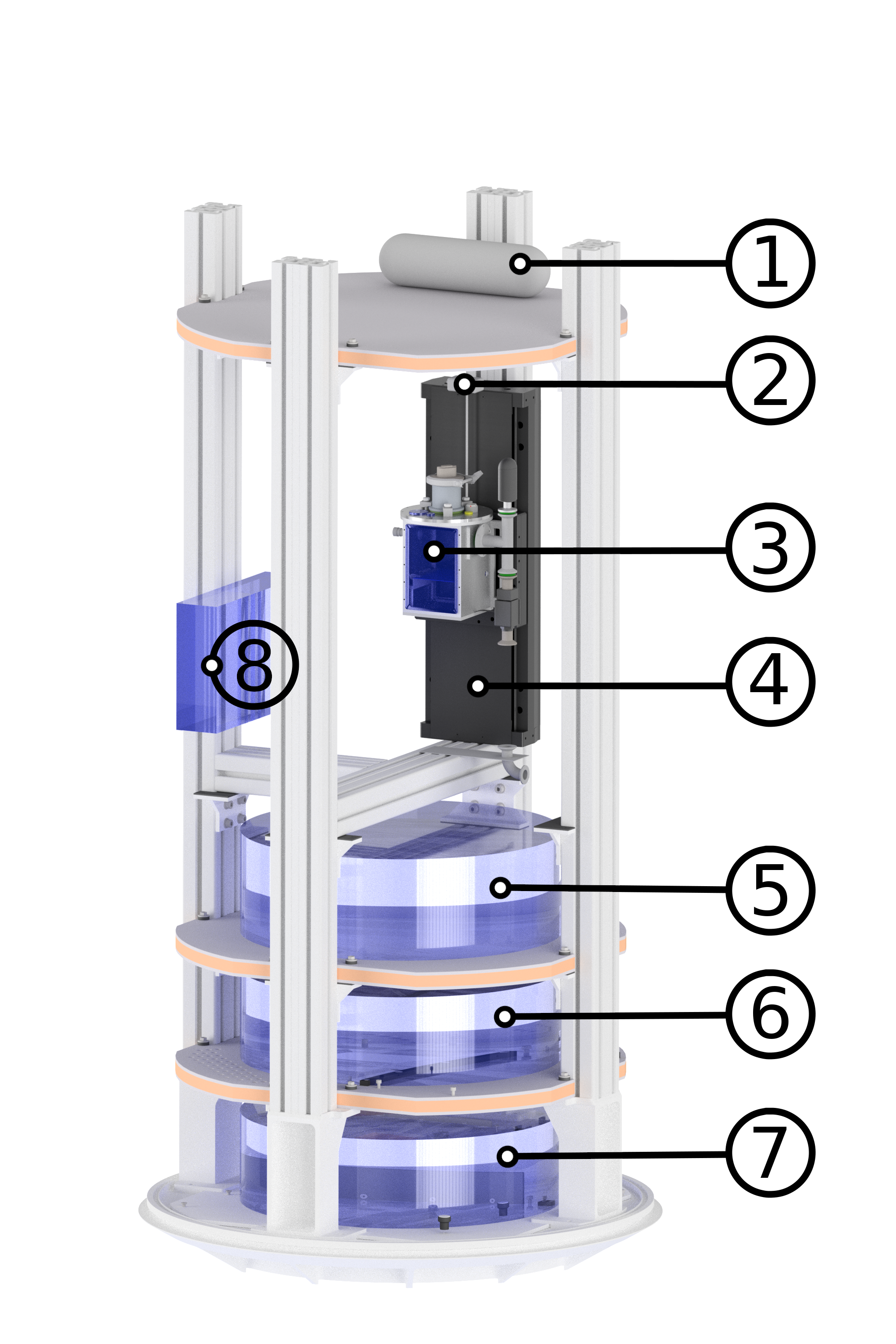}
    \caption{Full setup including drop tower capsule. (1): Pressure reservoir, (2): Brake, (3): Vacuum chamber, (4): Linear Stage, (5): Upper control level, (6): Lower control level, (7): Battery level, (8): Inverter}
    \label{fig:capsule}
\end{figure}

A bare drop capsule only containing batteries and basic capsule control hardware as detailed in \citep{dropmanual} was provided by ZARM. Our system was then arranged inside the capsule as is shown in Fig. \ref{fig:capsule}. 
%The exact configuration underwent several changes during development, but the main purpose and general layout remained mainly unchanged. That said, Fig. \ref{fig:capsule} is a schematic snapshot from late 2022. \\ \\
The central upper part of the capsule is inhabited by a Newport IMS-LM300 linear stage (4) carrying the vacuum chamber (3).  In this part of the capsule the most free space is available, which is needed for the moving stage and chamber to reach maximum travel and thus maximum milligravity time. Lateral dimensions are indicated in Fig. \ref{fig:platte} (a). The stage and chamber are oriented in vertical direction, which is the flight direction of the capsule. This is the only direction in which the stage fits the capsule. It also has the advantage that the direction of gravity will not change between ground and flight.  In our case that means the direction of gravity always remains perpendicular to the simulated asteroid surface, which eases its retainment during the rapid acceleration phase of the catapult launch. Attached to the vaccum chamber is a braking system, see Fig. \ref{fig:capsule} (2). The brake dampens the stage's and vacuum chamber's movement during launch and landing of the drop tower capsule. The brake is pneumatically operated and requires a pressure reservoir, see Fig. \ref{fig:capsule} (1), on top of the capsule. Another part of the setup integrated in the upper part of the capsule is the inverter (8). The inverter is necessary to drive the linear stage, which can't work with the direct current provided by the capsules system. 
%The trade off is that this may introduce disturbances for sensitive measurement, which however in most cases does not affect us. 
Just below the vacuum chamber and linear stage two control levels are located, at Fig.  \ref{fig:capsule} (5+6).
The upper control level (5) houses the Newport XPS-RLD controller for the linear stage. It provides a web interface to program and control the stage's movements, reads its encoder and provides power. It is accessed via ethernet with a passively cooled, automotive rated computer on the same level, which in turn can be reached over a remote desktop connection from outside the capsule. The upper control level furthermore contains some low voltage regulators and controls for details of the experiment.
The lower control level, Fig. \ref{fig:capsule} (6),  exclusively contains ZARM capsule and experiment control equipment. A PXI by National Instruments records experiment parameters and provides a plethora of digital and analog IO capabilities. %Recently the system changed slightly and is now running a full Linux OS. 
Below the control levels, at Fig. \ref{fig:capsule} (7), the power distribution unit (PDU)is situated. During flight, the capsule needs to be autonomous, so the PDU contains battery packs. The batteries are sufficient to supply $40\,$A at $24\,$ V on six controllable channels. % If different voltages are needed, $DC-DC$ or $DC-AC$ conversion needs to be done. 
This capsule is also compatible with the GTB Bremen Pro, offering lower microgravity time at a higher repetition rate. Thus, our capsule, an in turn our controlled milligravity platform  and our experiment can be used in either carrier.
\subsection{Stage and controller}
\begin{figure}
    \centering
    \includegraphics[width=0.95\linewidth]{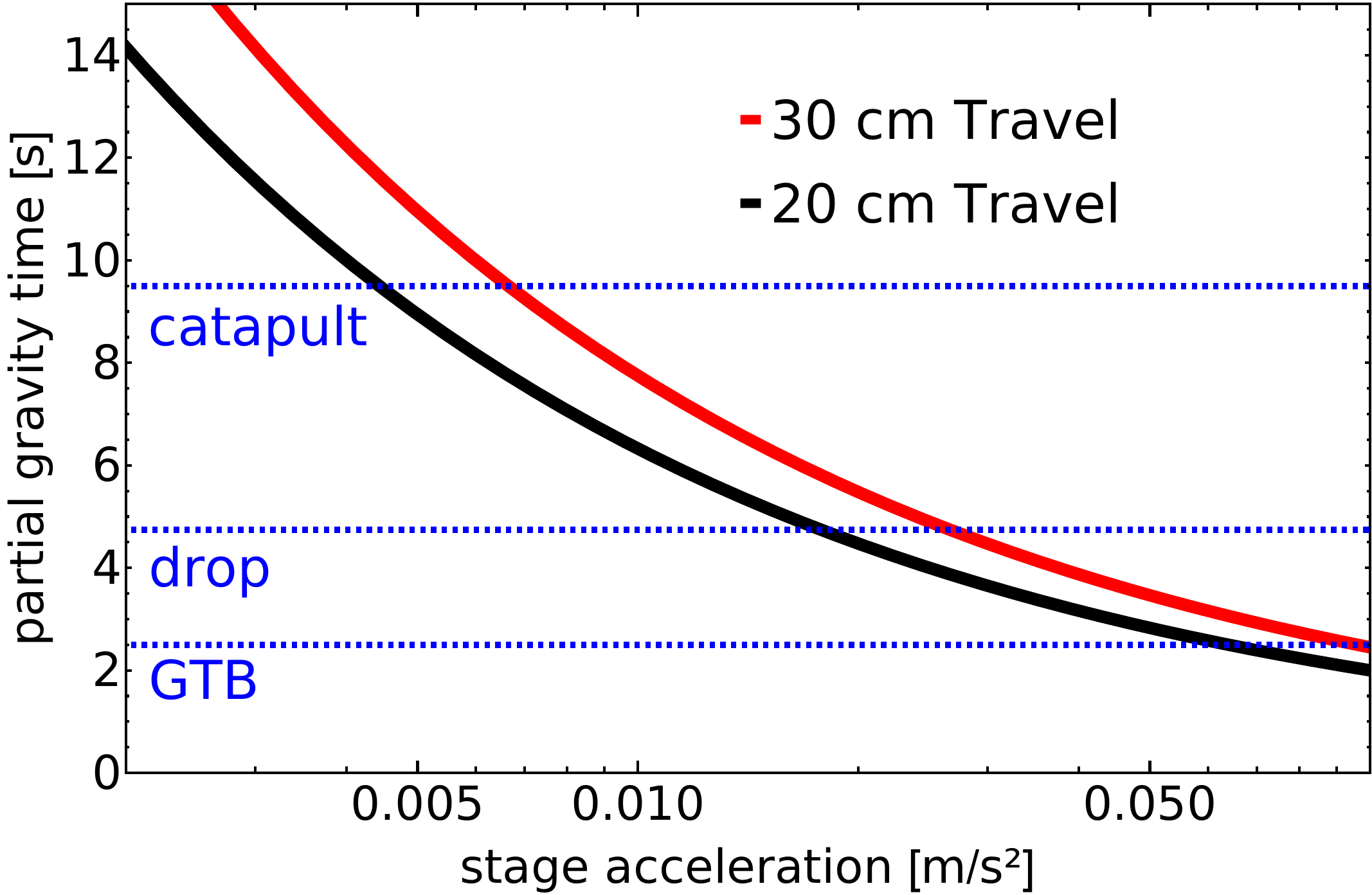}
    \caption{Possible experiment duration depending on partial g level with different travel distances and example carrier platforms. The blue curve denotes a configuration with $20\,$cm possible travel distance, the red curve the $30\,$cm. Duration cutoffs from example carrier platforms are indicated with blue dashed lines with the Gravitower Bremben Pro (GTB), the ZARM tower in drop configuration (drop) and the tower in catapult mode (catapult).}
    \label{fig:travele}
\end{figure}

The linear stage,
%in the final iteration of the experiment 
is a Newport M-IMS300LM-S \citep{newport-stage} brushless 3-phase direct drive. This component, serving as the partial gravity generator converting microgravity to partial gravity, is the core of our partial g platform. To fit inside the drop tower capsule and fully utilize the available travel distance, we chose the $300\,$mm travel model. This motorized stage is marketed by the manufacturer for precision optics, with a minimum incremental motion of $20\,$nm and a typical bidirectional repeatability of positioning of $\pm 1.7 \, \mu$m and a typical yaw of $\pm 25 \, \mu$rad. This precision and the accompanying mechanical rigidity is necessary for our application case to achieve smooth and controlled motion even at very low velocities and low accelerations.
%Mechanical rigidity for us is as important as precise control. 
The acceleration and deceleration phase of a catapult launch impose a large mechanical load on the stage. At accelerations of up to $50\,$g, an experiment of $2\,$kg or more becomes equivalent to a load of $\approx 1000\,$N. While in planned operation, this full load is never put on the stage and instead returned to a safe rest, see section \ref{sec:oper}, the system still needs to be designed to be as sturdy as possible. With that in mind, we chose the M-IMS300LM-S with a high central load capacity of $600\,$N. The experiment is bolted to the sled using $4$ M6 bolts. The stage has a top speed of $1000\,$mm\/s and a maximum acceleration of $40\,$m/s$^2$. As controller we use the Newport XPS-RLD \citep{newport-controller} with XPS-DRV11 motion controller card. The motion controller card is able to supply $600\,$W and $6\,$A peak and $300\,$W at $3\,$A continuous per channel, but is limited by the compact controller to $300\,$W max. Controller, card and stage operate in a closed loop, monitoring and correcting the positioning. With a servo rate of $10\,$kHz, this system is able to change its velocity at a high rate, which is necessary for a clean acceleration. The controller provides several GPIO channels, which can be conveniently used for trigger in- and outputs. For control, configuration and programming the controller provides a web interface accessible via ethernet. Software wise, the stage is controlled using a graphical user interface or a scripting language. For the experiments detailed here we use the scripting interface and the function \emph{pt-trajectory}, where incremental  position and time are specified, while the controller tries to keep the acceleration constant between increments. Importantly, for trajectories programmed this way acceleration does not need to be constant. Fig. \ref{fig:travele} illustrates the available experiment time depending on the chosen level of acceleration. Stronger acceleration of the linear stage's sled translates to a higher partial gravity level but also consumes the available travel faster, with the result that the acceleration can be maintained for a shorter time. Blue dashed lines in Fig. \ref{fig:travele} indicate the intrinsically given time limits imposed by each available carrier, with $9.5\,$s seconds for the ZARM drop tower in catapult- and $4.75\,$s in drop mode, as well as the Gravitower Bremen Pro (GTB) with $2.5\,$s. 
\subsection{Vacuum chamber}
The vacuum chamber is shown in detail in Fig. \ref{fig:chamber}. The top part Fig. \ref{fig:chamber} (1) is detailed below in Fig. \ref{fig:launcher}. As mentioned above, two cameras are mounted on top of the chamber. For them to operate, windows, see Fig. \ref{fig:chamber} and \ref{fig:cams}, are needed as viewports and for illumination. Illumination is provided by high power LEDs through the top viewports and / or by a backlight panel. At the sides of the chamber, Fig. \ref{fig:chamber} (3), the vacuum ports are located. Here, pressure sensors, additional valves to ventilate the chamber and the magnetic valve mentioned in Fig. \ref{fig:vac} (4) are connected. The main windows, Fig. \ref{fig:chamber} (4), cover the whole front- and backside of the vacuum chamber. They enable illumination and observation of the whole chamber volume. The mechanism at Fig. \ref{fig:chamber} (5) is modular and can be fully removed from the chambers interior. It is used to cover the granular bed at Fig. \ref{fig:chamber} (6) until shortly after the acceleration phase, preventing the granular specimen to escape into the whole chamber volume, thus forcing it to settle more quickly.
\begin{figure}
    \centering
    \includegraphics[width=0.9\linewidth]{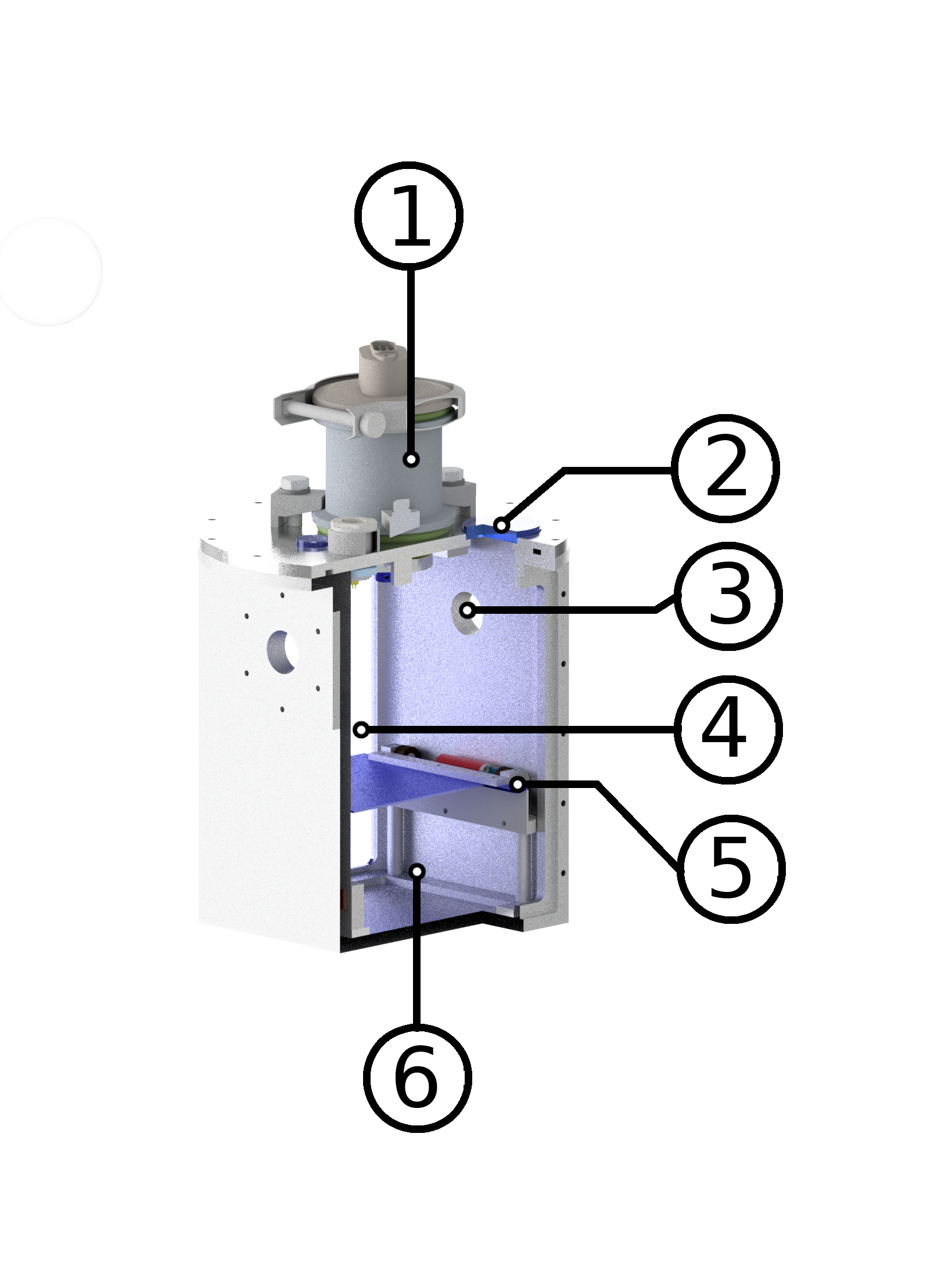}
    \caption{Vacuum chamber. (1): Launcher compartment, (2): Top windows, (3): Vacuum Ports, (4): Main viewport, (5): Cover, (6): Granular bed compartment}
    \label{fig:chamber}
\end{figure}
\subsection{Launcher}
\begin{figure}
    \centering
    \includegraphics[width=0.8\linewidth]{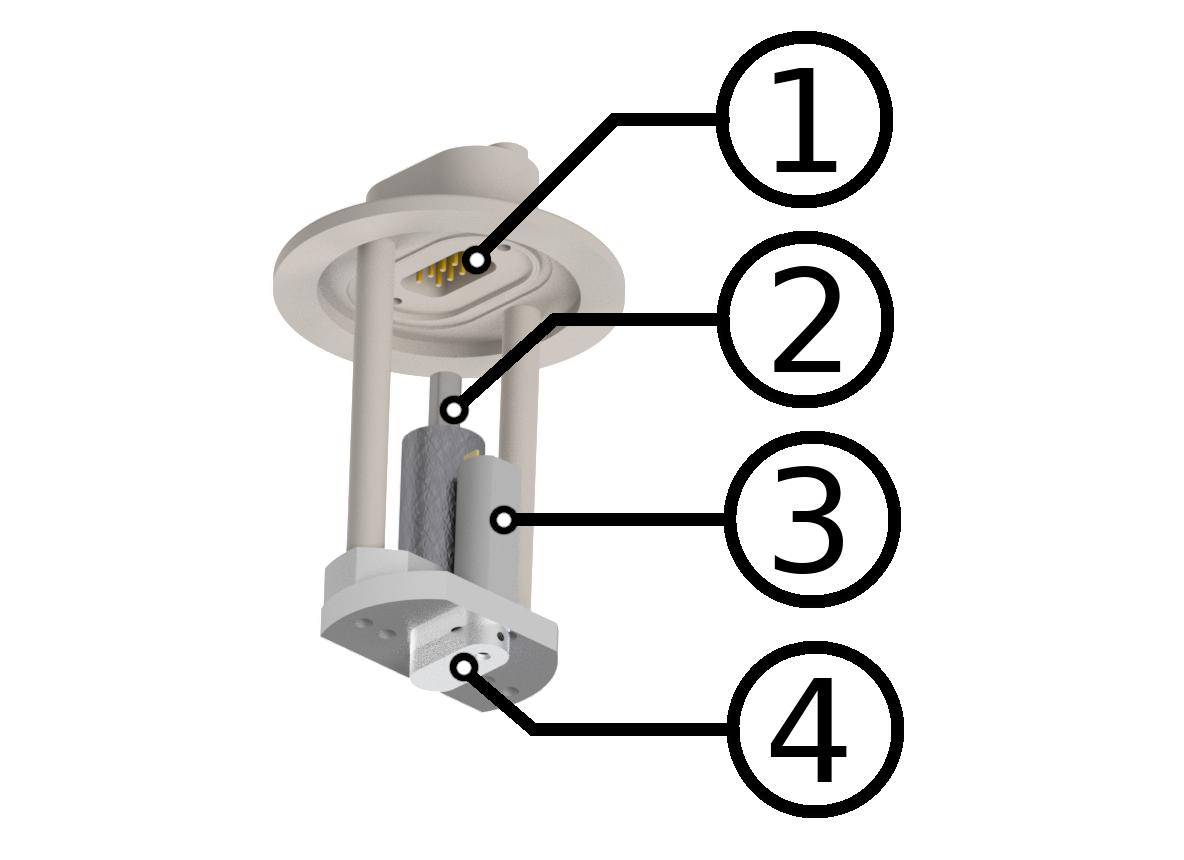}
    \caption{Launcher. (1): Electric feedtrough, (2): DC Motor, (3): Launcher Magnet, (4): Impactor hatch}
    \label{fig:launcher}
\end{figure}

For a schematic of the launcher, please view Fig. \ref{fig:chamber} (1) and Fig. \ref{fig:launcher}. The launcher is mounted at the top of the vacuum chamber to shoot impactors from a perpendicular angle onto the granular bed. The launcher consists of a ISO-KF 40 flange with D-Sub electrical feedtrough as structural base. Mounted to that flange is the launching mechanism itself, consisting of a hatch (4) reatining the impactor, a magnet (3) and a motor (2). Once the motor (2) has rotated the hatch to release the impactor, the magnet (3) is powered to punch a metal rod through a hole above the impactor, pushing it down and towards the granular bed.  The duration and strength of the voltage applied to the magnet sets the impactors initial speed. One great advantage of this system is the extremely low minimal impactor velocity. Lowest tested impact speeds range down below $4\,$cm\/$s^2$. The mounting braces connecting sample holder and flange can be extended to reduce the time of flight before the impactor reaches the granular bed. The whole assembly can be exchanged for different probing mechanisms if necessary. 
\subsection{Cameras and Lighting}
\begin{figure}[ht]
    \centering
    \includegraphics[width=0.8\linewidth]{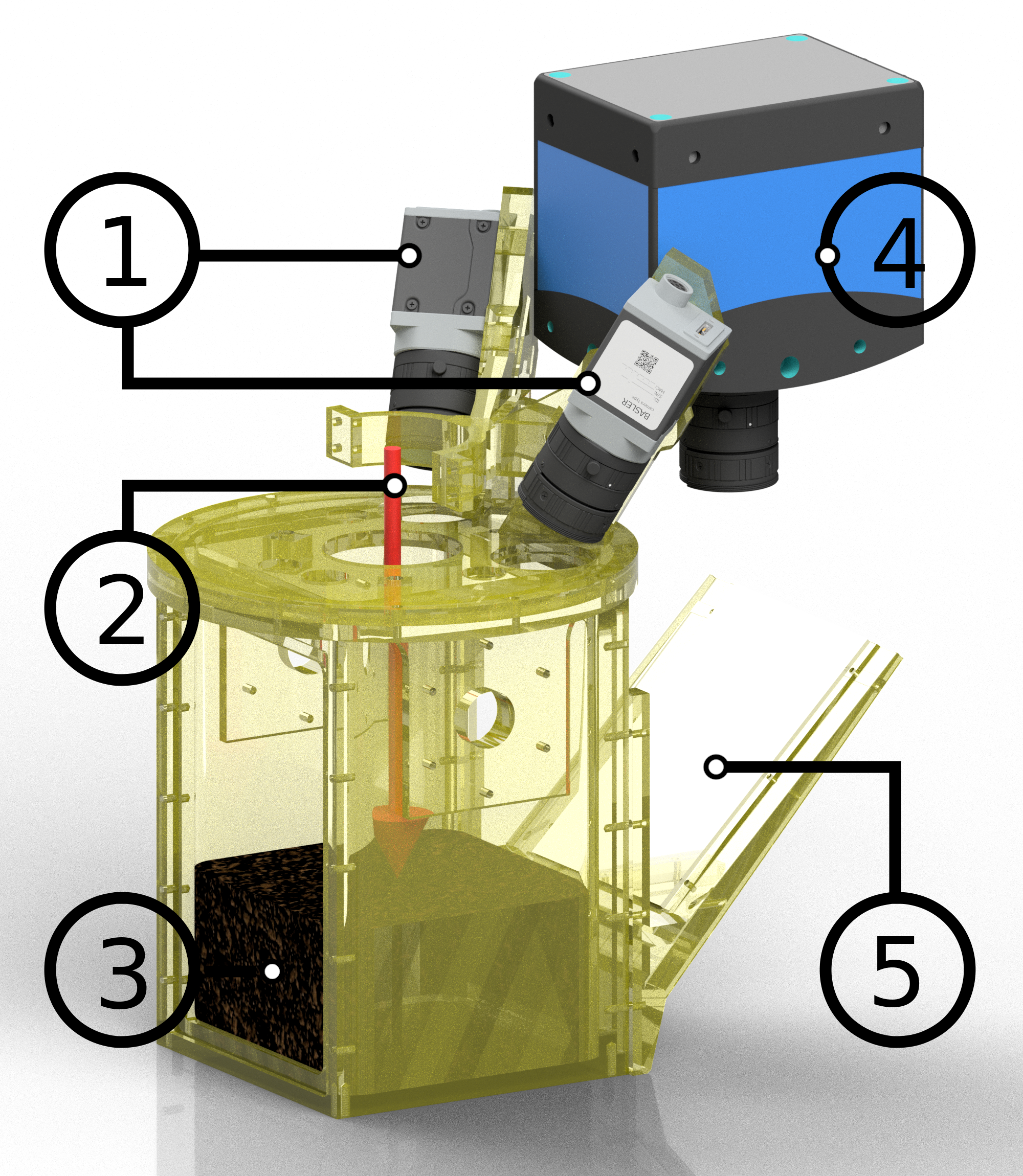}
    \caption{Camera setup. (1): Top cameras, (2): Impactor trajectory, (3): Granular bed, (4): Main camera, (5): (Mirror)}
    \label{fig:cams}
\end{figure}

The optics setup is shown in Fig. \ref{fig:cams}. The vacuum chamber is depicted in transparent yellow, with the granular bed Fig. \ref{fig:cams} (3) at the bottom. The impactor's approximate trajectory is indicated by a red arrow Fig. \ref{fig:cams} (2) pointing from the launcher, which is omitted in Fig. \ref{fig:cams} for clarity, down onto the granular bed. The two cameras Fig. \ref{fig:cams} (1) are positioned above the bed, enabling an overview of the bed surface and possible non-normal movement components of the impactor before and after contact with the bed. Using two identical is a prerequisite for reconstruction of depth information. In Fig. \ref{fig:cams}, those top cameras are Basler a2A2590-60ucBAS USB3.0 cameras with a maximum resolution of $2592 \times 1944$ Pixels at $60\,$fps, equipped with 4mm lenses. As main camera we mount a Mikrotron Cube7 with a maximum resolution of $1696 \times1710$ Pixels at $528\,$fps at full reslution equipped with a $16\,$mm lens. The measured spatial resolution at the vacuum chamber center results to $\approx 90\,$mm$/1326 \,$px$=68\,\mu$m$/$px. The Mikrotron camera houses enough internal memory to store a full sequence recorded in the drop tower and a is equipped with a backup battery. Using the internal storage reduces the bandwidth needed to load image data in real time onto the computer which controls the other cameras and the linear stage. % For GTB operation this is less convenient, as the full sequence needs to be downloaded after each parabola.

\subsection{Pump Loop}
\begin{figure}
    \centering
    \includegraphics[width=0.9\linewidth]{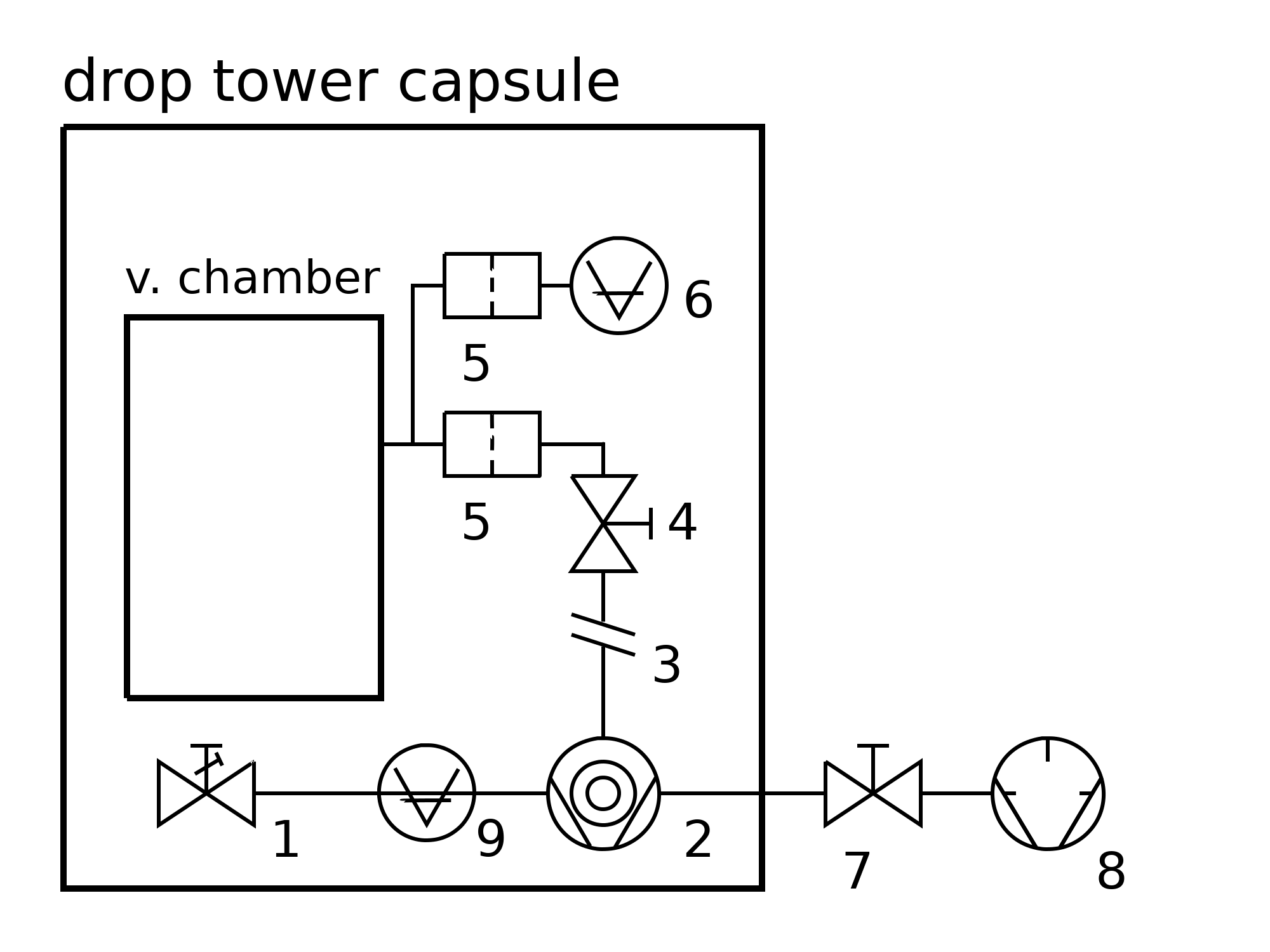}
    \caption{Vacuum setup. (1): Needle valve, (2): Turbomolecular pump, (3): Disconnection, (4): Magnetic valve, (5): Filter, (6): Chamber pressure sensor, (7): Magnetic valve, (8): Rotary vane pump,  (9): Vent-line pressure sensor}
    \label{fig:vac}
\end{figure}

The drop tower is evacuated to $<0.1\,$hPa during launch, but the capsule pressure is kept at $10^5\,$hPa at all times. Hence the vacuum chamber, needs to be evacuated. This requires integration of a pumping system into the capsule. A schematic portrait of the pumping loop can be seen in Fig. \ref{fig:vac}. Directly attached to the vacuum chamber is a pressure sensor (Leybold Thermovac TTR 91 R) Fig. \ref{fig:vac} (6), behind a filter (5) protecting it from contamination with dust. This is used to monitor the chamber's pressure during all times. At Fig. \ref{fig:vac} (4) a magnetic valve (Pfeiffer DVI 005 M) is mounted to the chamber, like the sensor it is protected by a filter (5). % The reason to use two separate filters is for constructive reasons obscured by the simplicity of this sketch.
The mangetic valve (4) can be shut to disconnect the vacuum chamber from the rest of this loop at point (3) to be able to move the vacuum chamber and stage without any possible disturbance of a vacuum hose. At the other side of point (3), the  'vent-line' side, still inside of the drop tower capsule, a turbomolecular pump (XXTypeXX) (2) is installed. It is connected to a sensor (9), monitoring the pressure of the vent-line side and a needle valve (1). Sensors (9) and (6) are used in combination to ensure that the pressure differential between vent-line and vacuum chamber is below a few hPa, to avoid an abrupt pressure increase in the chamber when opening valve (4). Furthermore, a pressure of $<10\,$hPa at sensor (9) is the criterion for a safe start-up of the pump (2). To stop the pump (2) within a few seconds before the catapult launch, the needle valve (1) can be opened. Outside of the drop tower capsule, an aditional magnetic valve (7) is located to seal the whole vacuum system towards the outside. The rotary pump (8) is employed to provide a low enough pressure inside the vent-line for the turbo-pump (2) to operate, while the capsule is located outside the drop tower. It is disconnected and valve (7) is shut when the capsule is placed inside of the drop tower. Once the ammbient pressure inside of the drop tower reaches the turbo-pump's operating pressure, the valve (7) can be opened and the pump (2) started.
\section{Operation\label{sec:oper}}
\begin{figure}
    \centering
    \includegraphics[width=0.95\linewidth]{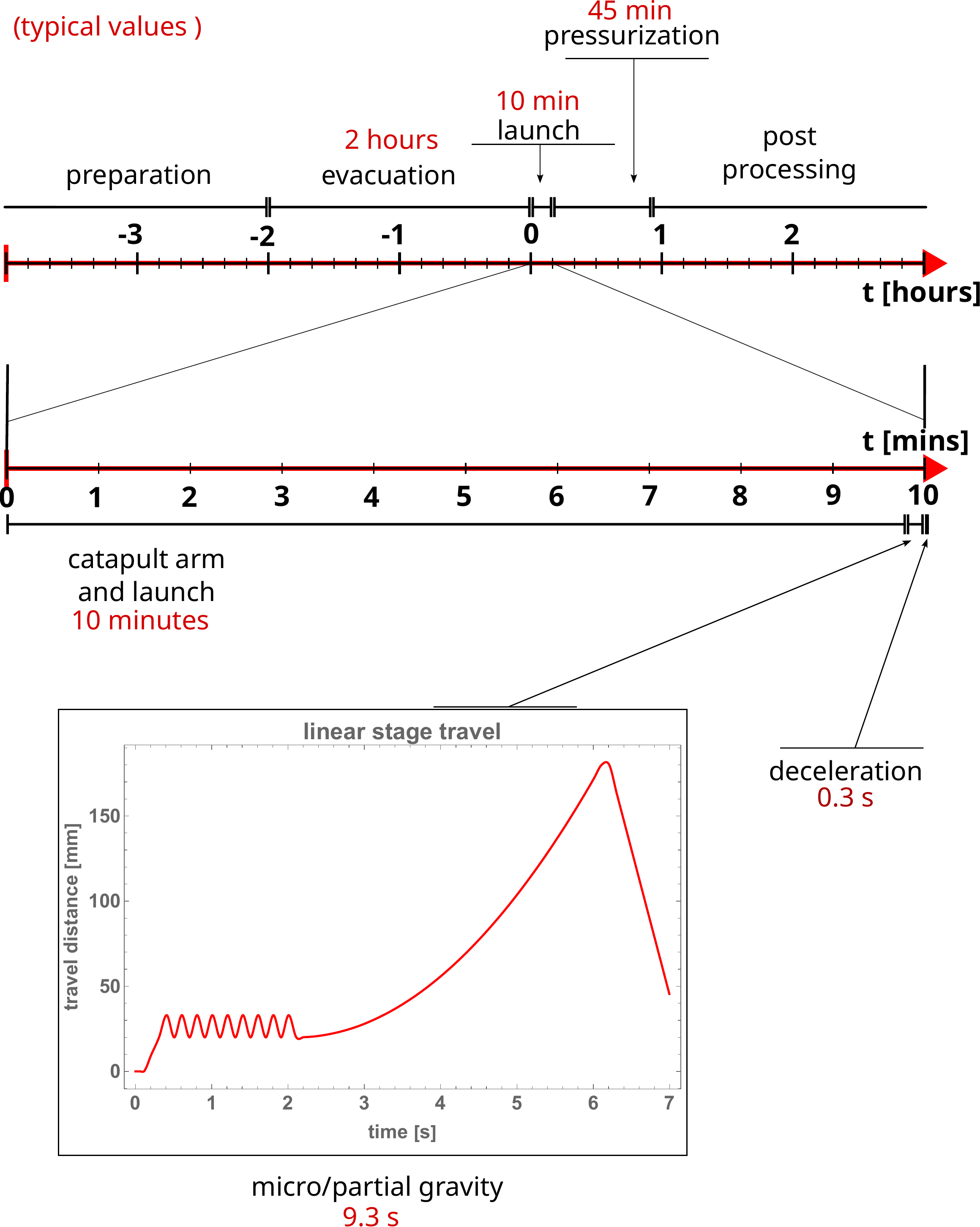}
    \caption{Sketch of timeline of catapult launch, not to scale. Top line: Time scale of hours, repeatable up to two times per day. Bottom line: Launch sequence, timescale in Minutes. Inset: Flight, timescale in seconds.}
    \label{fig:timeline}
\end{figure}

Operation of the experiment is structured in five different top level phases. Each phase is detailed below and a Timeline of all phases is shown in Fig. \ref{fig:timeline}. \textbf{1. Preparation:} The experiment is prepared outside of the drop towers vacuum tube. In case of the asteroid surface experiment the impactor is loaded into its launcher, the vacuum chamber is cleaned and then filled with the regolith simulant. The pressure reservoir is pressurized. All adjustments to the cameras have to be completed in this phase, as they cannot be physically accessed later on. The capsule is disconnected from external power, data connections, external pump and pressure reservoir, sealed and positioned on the catapult inside of the drop tower.  \textbf{2. Evacuation:} The drop tower's vacuum tube is evacuated. During this phase, the experiment is monitored and controlled via remote desktop connection. Valves and pumps are operated using the drop tower capsules integrated control system. In this phase, via remote control with the camera computer, the cameras are armed to be ready to accept a hardware trigger signal synchronized to the catapult launch. With experiment lighting turned on, this is the last possible occasion to check camera parameters. When the drop tower pressure is lower than the vacuum chambers pressure, the ventline valve and vacuum chamber valve are opened. The turbomolecular pump is started. As soon as the catapult is ready to fire, the turbomolecular pump is powered off, magnetic valves shut and the pump is stopped by increasing the pressure using the needle valve. As soon as the turbopump is at a halt, the capsules launch can be manually triggered. \textbf{3. Microgravity:} During microgravity, external control is not possible. A predefined control sequence is executed by the capsule computer and linear stage controller. After a preset delay after launch, the capsule computer sends trigger signals to cameras and linear stage and the break is released. The cameras start recording and the linear stage initializes, getting ready for motion. After initialization the the stage starts executing the programmed trajectory. In case of the asteroid surface experiment, to ensure a defined surface of the regolith layer, the acceleration is set to start with a higher value than the targeted asteroid gravity level for $\approx0.3\,$s. This initial higher gravity level allows for disturbances in the regolith caused by the catapult launch to settle. After that, the stage will accelerate at the desired partial gravity level. For the asteroid experiment, once the desired partial gravity level has been reached the hatch covering the regolith bed is opened and the impactor is launched whith the camera system recording the interation beween impactor and regolith bed. Shortly before landing, that is, in time for deceleration of the capsule, the linear stage moves the experiment chamber into its safe starting position to avoid mechanical stress, and the peneumatic break is engaged again. The capsule then impacts the deceleration container that has moved to its position right above the catapult and comes to rest inside a bed of Styrofoam granules. \textbf{4. Pressurization:} Before the experiment can be physically accessed again, the drop towers vacuum tube needs to be pressurized. This takes about $45\,$Minutes. \textbf{5. Post-Launch routine} As soon as the tower has reached ambient pressure, the capsule is recovered from the deceleration container and moved to the experiment bay, where it is connected to all necessary external resources. Here, the data acquired during flight can be backed up to external discs and a first ad-hoc evaluation of the experiment can be performed. The system is connected to external vacuum lines, to enable controlled pressurization of the vacuum chamber. The experiment can now be prepared for next launch.

\section{Perfomance}
\begin{figure}
    \centering
    \includegraphics[width=0.5\textwidth]{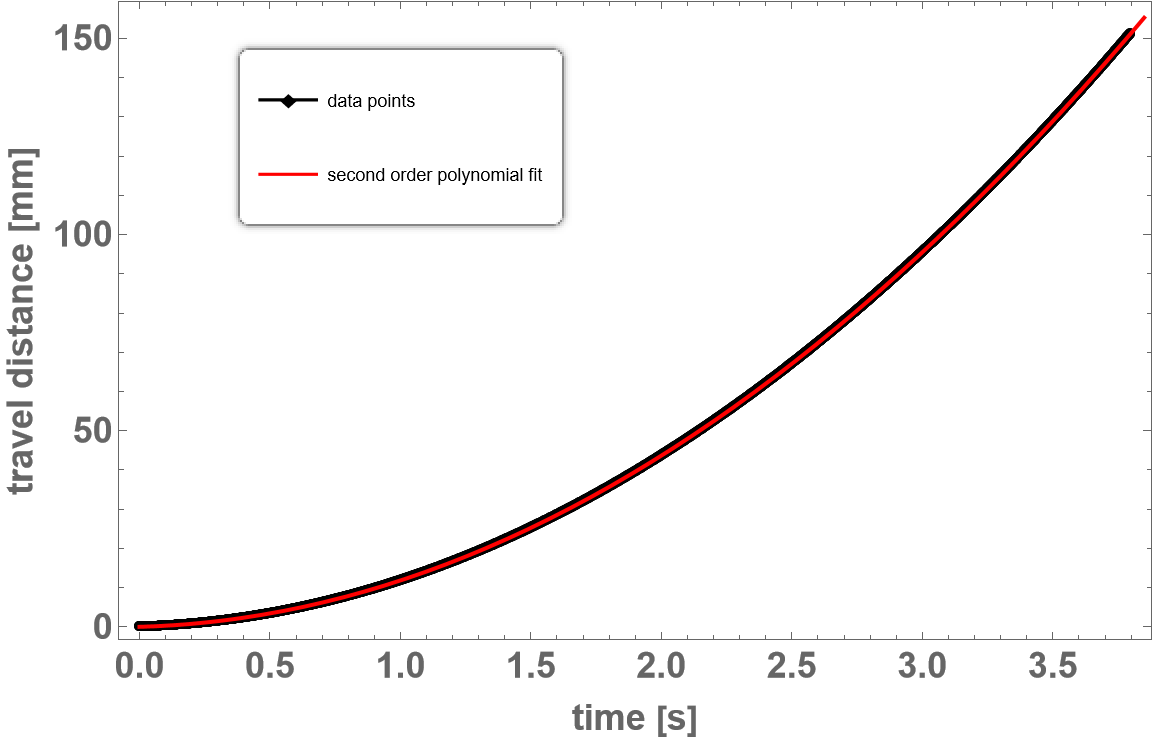}
    \caption{Linear stage stage trajectory (black) with fitted polynomial (red) $x(t)= 0.5\cdot a \cdot t^2 + v_0 \cdot t + x_0$ with $a=20 \,$mm\/s$^2$, $v_0=1.81\,$mm\/s, $x_0=0.0522\,$mm}
    \label{fig:stageperf}
\end{figure}
\begin{figure}
    \centering
    \includegraphics[width=0.5\textwidth]{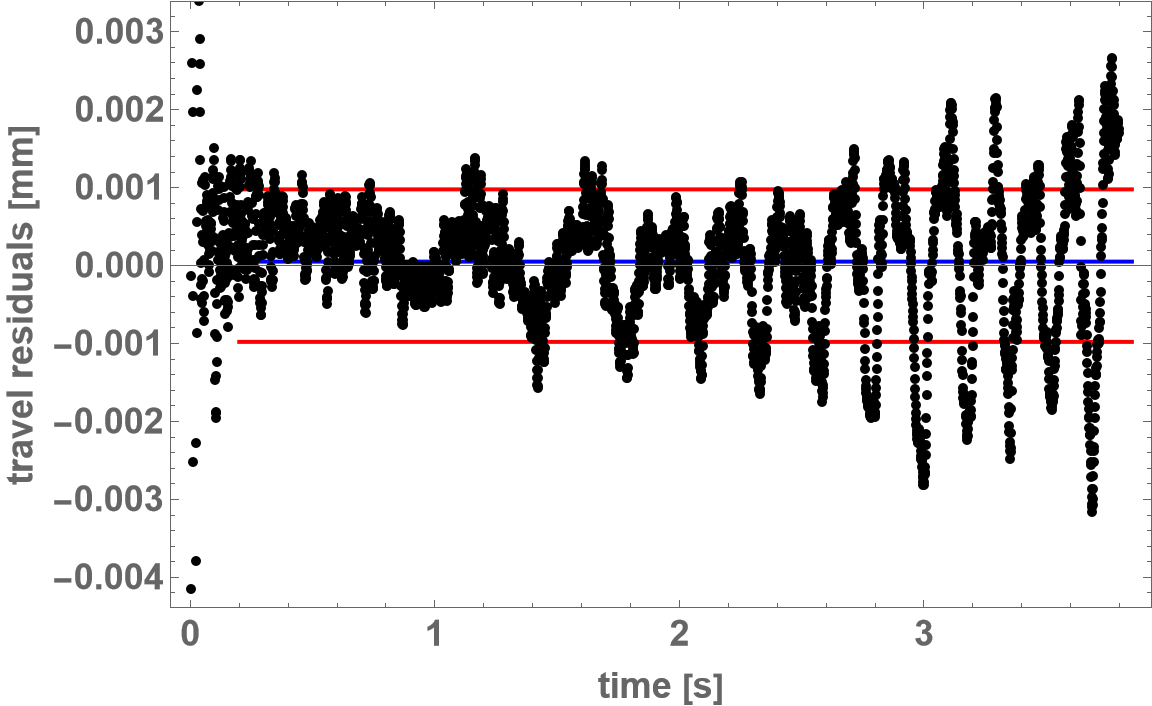}
    \caption{Residuals of fit to stage trajectory. Mean ($ -5 \cdot 10^{-5}\,$ mm) in blue, standard deviation ($9.78\cdot 10^{-4}\,$mm) in red.}
    \label{fig:stageres}
\end{figure}

To evaluate the milligravity performance of our system we used three different kinds of measurement. % at different times of the experiment's lifetime. 
With a magnetic measurement tape and a hall sensor we recorded the position of the linear stage during our first campaigns, because the linear stage used at that time (Thorlabs DDS220/M) was not able to output the position at a high time resolution. During later campaigns, with the Newport M-IMS300LM-S the integrated measurement system could be used for output. Additionally, tracking the ballistic arcs of free flying particles inside the experiment chamber provides a universally applicable and precise way to estimate gravity levels. The data gathered by the stage-controller system is output in $ms$  and $nm$ increments. A parabola, described by $x(t) = 0.5 a t^2 + v_0 t +x_0$  is used to fit the tractory, as shown in Fig \ref{fig:stageperf}. Here, the data points are shown in black, with the fit in red with $x(t)= 0.5 \cdot a \cdot t^2 + v_0 \cdot t + x_0$ with $a=20 \,$mm\/s$^2$, $v_0=1.81\,$mm\/s, $x_0=0.0522\,$mm. The fit aligns with the data almost perfectly, no deviation can be seen by eye. Consequentially, the residuals, i.e. the differences between fit and measured data are shown in Fig.\ref{fig:stageres} for each time point. After a short period of equilibration, roughly $0.2\,$s, the difference between measured position and a perfect parabolic trajectory are less than $3.16 \cdot 10^{-3}\,$mm. The mean deviation lies at $-5 \cdot 10^{-5}\,$mm, marked in blue, the standard deviation at $9.78\cdot 10^{-4}\,$mm, marked with a red line. Typical time scales for deviations from mean are between $0.5\,$s and $0.1\,$s for this trajectory. With a typical error in the range of micrometers, the performance of our system meets all our requirements and is sufficient for even the most delicate granular experiments. The performance at lower linear stage acceleration, with $a=2\,$mm/s$^2$ is evaluated in \citet{Joeris} and shows comparably low errors. Tested and possible trajectories do not only include constant accelerations for partial gravity, but also over-acceleration for specific times as outlined in \citet{Joeris} or  vibration as hinted in Fig. \ref{fig:timeline}.
For the asteroid experiment, the launcher has proven to be reliable and controllable. With the hatch safely retaining the impactors during catapult launch, it is designed for use under extreme high gravity conditions. Not only is it able to store and launch spherical particles with a diameter of $3\,$mm, but it can handle irregularly shaped objects as well. There is a limit to that though, as impactors with extreme geometries may get stuck, so test-launching impactors is advised. 
For use in the Gravitower Bremen Pro, a high repetition rate is desirable. Our combination of linear stage and vacuum chamber achieved a maximum of $15$ repetitions per half-day, demonstrating the possibility to produce significantly more data points than by using the big ZARM drop tower in catapult mode. The limiting factor however is not the linear stage, but the combination of launcher and vacuum chamber, which need to be accessed after each flight and the main camera, from which data download is possible only via gigabit ethernet. Without this limitation and for other experiments that do not require reloading the launcher or using this specific camera, the repetition rate of our platform including the linear stage may be significantly increased.

\section{Science and Data}
\begin{figure}[ht]
    \centering
    \includegraphics[width=0.95\linewidth]{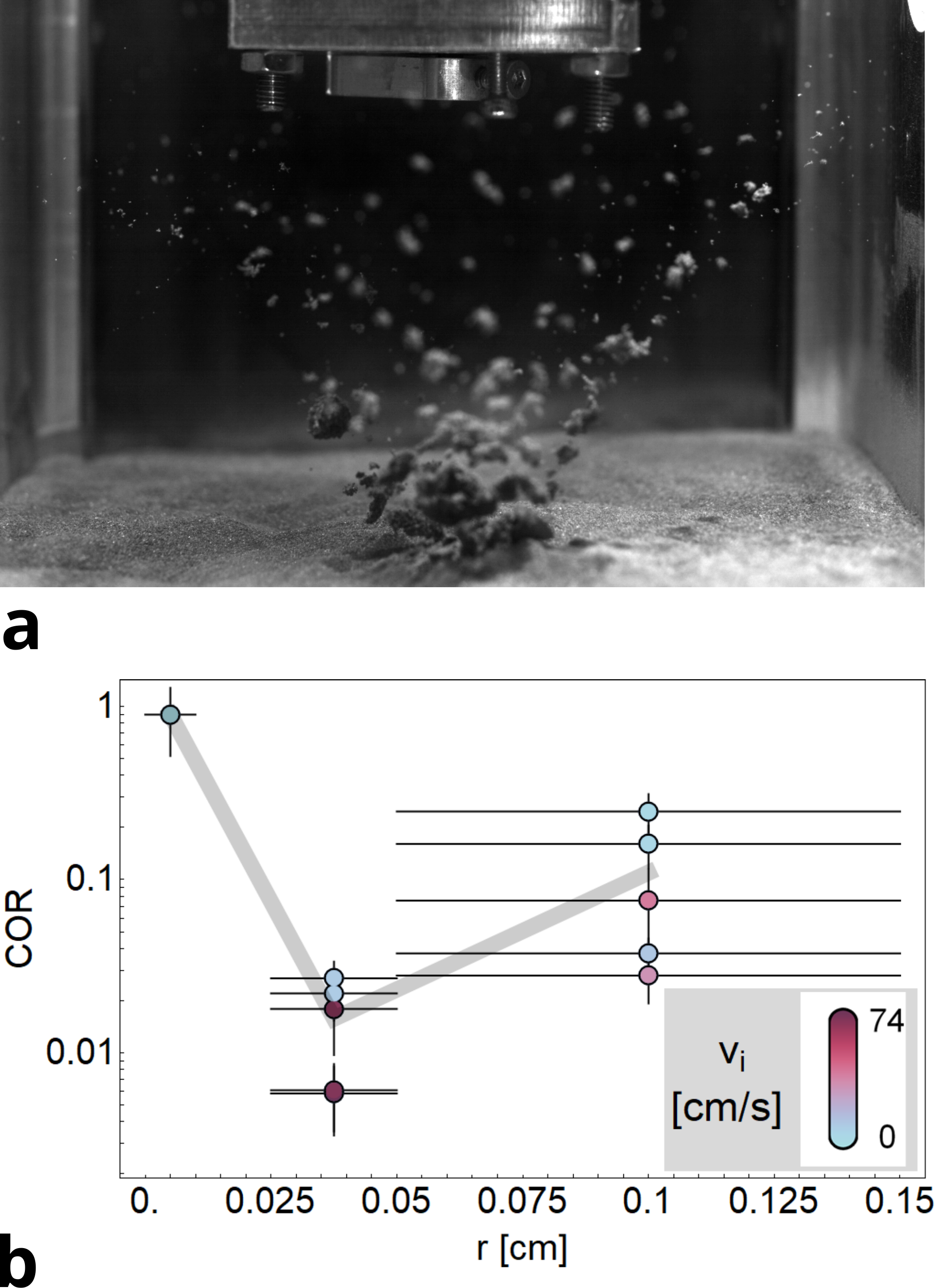}
    \caption{Experiment results. a) Impact splash recorded with high speed camera under microgravity. b) Coefficient of resitution over target particle size. Image taken from \textit{ Joeris, K., Schönau, L., Keulen, M. et al.  npj Microgravity 8, 36 (2022)}, licensed under a Creative Commons Attribution (CC BY 4.0) license \citep{joeris2022influence}.}
    \label{fig:exres}
\end{figure}

The controlled milligravity platform has proven itself in several measurment campaigns, producing scientific output. As in \citet{joeris2022influence}, it was used to generate impacts under asteroid conditions and investigate the impactor kinetics. The coefficient of restitution was measured under milligravity and compared to simulations. It was found, that the coefficient of restitution of a granular bed in low gravity does not behave monotonically with respect to bed particle size. Instead, a strong minimum was found for intermediate bed particles, slightly smaller than the impactor at a size of $0.025\,$cm to $0.5\,$cm. The comparison with simulations yields the conclusion, that cohesion in some form is a probable driver for this qualitative behaviour. This is points towards the necessity of a more complex sorting kinetic sorting theory than the ballistic sorting effect \citep{Shinbrot}. Furthermore, settling experiments were performed, investigating packing density of granular beds under reduced gravity. In this context, the platform demonstrated its capability to apply mechanical agitation (shaking) to the experiment and then drive at defined accelerations, without affecting its performance negatively.\\
Additionally, the ejecta generated during impacts, as shown in Fig. \ref{fig:exres} has been studied for varying impactor velocities. The clean partial gravity allows for exclusion of undesirable influences on the granular contacts, which may disturb the characteristics of ejecta generation. The dynamics studied here happen in the regime, where graviational and cohesive forces are of the same magnitude. This makes the granular surface susceptible for dusturbances of g-jitter. While providing a, to our knowledge, unprecedented quality of undisturbed granular surfaces at low partial gravity and extremely low velocity impacts, we are able to extend the data range of ejecta generation to low energy environments.
% bilder von daten rein, bild von splash

\section{Outlook/ Stage as Facility}
\begin{figure}[ht]
     \centering
     \includegraphics[width=0.95\linewidth]{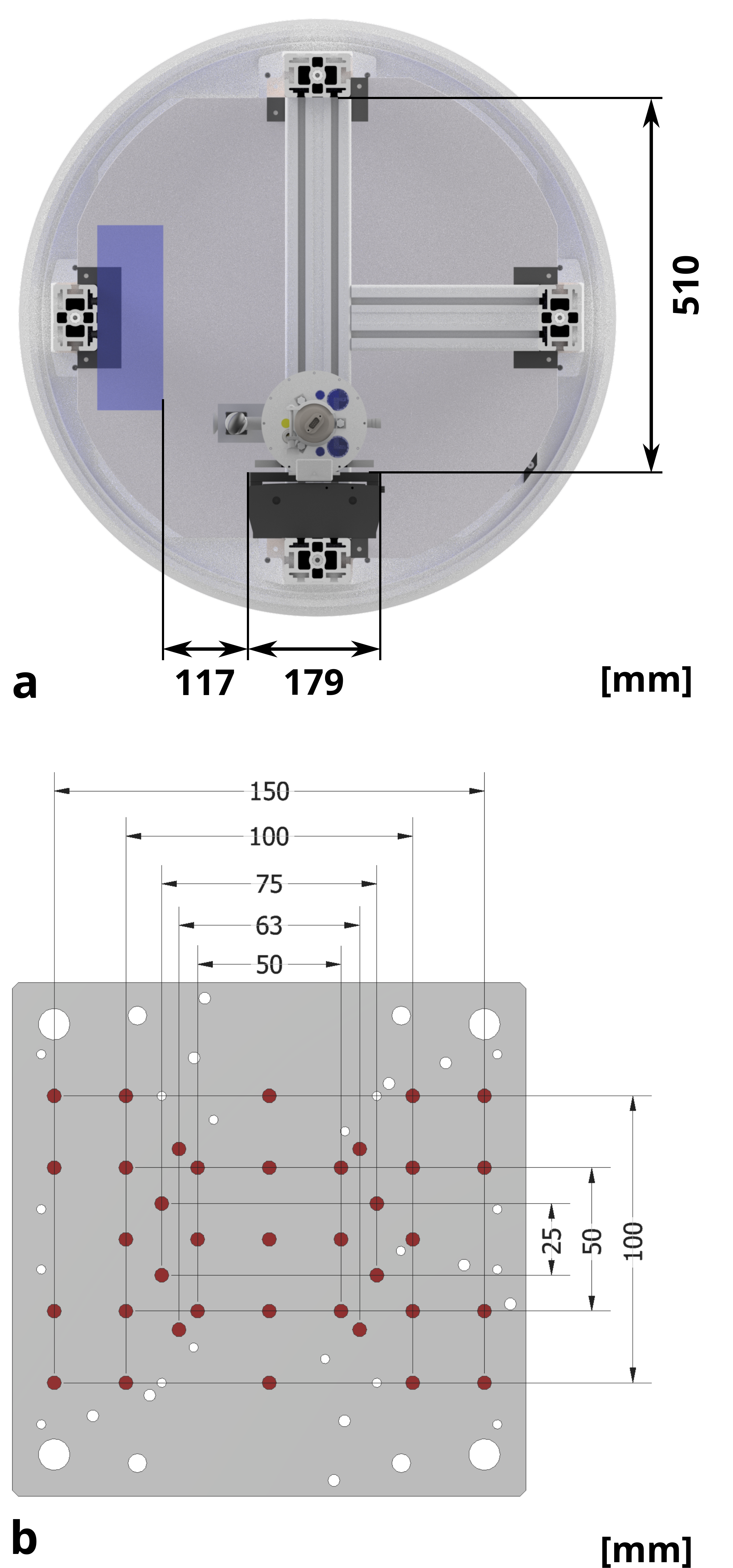}
     \caption{a) Topview cutaway of drop tower capsule. Annotations of Lateral dimensions available for experiments. b) Linear stage mounting plate. Dimensions are give for all relevant M6 holes. For a complete list of threaded holes see \citet{newport-stage}}
     \label{fig:platte}
 \end{figure}

We understand the assembly of stage and drop tower capsule as an open environment for partial g experiments. As such, we are open to reasonable requests from other pojects/groups which can benefit from our work. We offer a platform with precisely programmable partial gravity environments, excelling especially at ultra low accelerations in the milligravity regime, but not limited to that. The platform is able to provide a steady partial gravity environment at acceleration values which can be chosen by the experimenter. The near future of our platform is already set, with an external request and two in-house experiments currently being conducted. For available lateral dimensions inside of the drop tower capsule, see Fig. \ref{fig:platte} a. To attach a different experiment chamber to the linear stage a  mounting plate $179\,$mm wide, with the $117\,$mm of space towards the inverter is attached. For stability it is advisable to keep the mass centered on the mounting plate. The distance between mounting plate and the opposing capsule stringer is $510\,$mm. Objects mounted to the stage are advised to no use all of this space, as the center of mass needs to be as close to the mounting plate as possible. This means that experiment parts with relevant mass, excluding e.g. data cables, should not extend more than approximately $200\,$mm in this direction. Panel b of Fig. \ref{fig:platte} shows an image of the linear stage's mounting plate, with exact dimensions found at the manufacturer's data sheet \cite{newport-stage}. A plethora of mounting holes is available with metric dimensions, from M4 to M6. While the mounting needs to withstand some load during capsule deceleration, we advise towards usage of the M6 holes.
%Bilder von Adapterplatte, Bilder von Gesamtaufbau, clearance, links zu technischen ressourcen in die refs
 \section{Author declarations}
 \section{Conflict of Interests}
 The authors have no conflicts to disclose.
 \section{Author Contributions}
  \textbf{Kolja Joeris:} Investigation (lead); Resources (lead); Writing - orignial draft (lead); Visualization (lead); \textbf{Matthias Keulen:} Investigation (supporting); Writing - original draft (supporting); Visualization (supporting). \textbf{Jonathan E. Kollmer:} Conceptualization (lead); Supervision (lead) Project Administration (lead); Funding Acquisition (lead); Writing - original draft (supporting). 
\section{Acknowledgements}
This work was supported by the DLR Space Administration with funds provided by the Federal Ministry for Economic Affairs and Climate Action (BMWK) based on a decision of the German Federal Parliament under grant number 50WM1943, 50WK2270C and 50WM2243.
\bibliography{bib}

\end{document}